\documentclass[twocolumn]{svjour3}          % twocolumn
\smartqed  % flush right qed marks, e.g. at end of proof 

\usepackage{color}

\usepackage{booktabs} % For formal tables
\usepackage{times}
\usepackage{xcolor}
\usepackage{soul}
\usepackage[utf8]{inputenc}

\usepackage{hyperref}
\usepackage{url}

\usepackage{grffile}

  \usepackage{tikz }
  \usetikzlibrary{calc}

\usepackage{csquotes}
\usepackage{url}
\usepackage{balance}  % for  \balance command ON LAST PAGE  (only there!)

\usepackage{amssymb}
\usepackage{amsmath}
\usepackage{graphicx}
\usepackage{subfigure}
\usepackage[pdf]{pstricks}

\usepackage{amsthm}

\usepackage{color}

\newcommand{\ie}{\emph{i.e.}, }%\xspace}
\newcommand{\eg}{\emph{e.g.}, }%\xspace}

\usepackage[makeroom]{cancel}

\usepackage{algorithm}
\usepackage{algorithmic}
\usepackage{ifsym} % symbole enveloppe
\usepackage{booktabs}

\theoremstyle{definition}

\usepackage{footnote}
\makesavenoteenv{tabular}
\makesavenoteenv{table}

\usepackage{color, colortbl}
\definecolor{Gray}{gray}{0.9}

\usepackage{flushend}

\title{Adversarial frontier stitching for remote neural network watermarking}

\begin{document}

% Multiple author syntax (remove the single-author syntax above and the \iffalse ... \fi here)
\author{
Erwan {Le Merrer} \and
Patrick P\'erez \and
Gilles Tr\'edan
}

%% \author{First Author         \and
%%         Second Author %etc.
%% }


\institute{\textit{Contact author:} Erwan {Le Merrer}  \at
  IRISA/Inria, Campus de Beaulieu,
  35 576 Cesson Sévigné, France\\ 
  Tel.: +33299847213\\
              \email{erwan.le-merrer@inria.fr}           %  \\
           \and
           Patrick P\'erez \at
           valeo.ai,  Creteil, Île-de-France, France\\
           \email{patrick.perez@valeo.com}           %  \\
           \and
           Gilles Tr\'edan \at
           LAAS/CNRS, 7 avenue du Colonel Roche, 31031 Toulouse, France
           \email{gtredan@laas.fr}           %  \\
}

\date{}

\maketitle

\begin{abstract}
\newenvironment{inline}[2]{
    \begin{figure}[h!]
        \begin{shaded}
            \caption{#1}\label{#2}
                \vspace{10pt}%
}{
        \end{shaded}
    \end{figure}
}
  
% motivation
The state of the art performance of deep learning models comes at a high
cost for companies and institutions, due to 
the tedious data collection and the heavy processing requirements.
Recently, \cite{Uchida:2017,Nagai2018} proposed to watermark
convolutional neural networks for image classification, by embedding information into their weights.
While this is a clear progress
towards model protection, this technique solely allows for extracting
the watermark from a network that one \textit{accesses locally} and entirely.

Instead, we aim at allowing the extraction of the watermark from a neural network (or any other machine
learning model) that is operated \textit{remotely}, and available through a service API.
To this end, we propose to mark the model's action itself, 
tweaking slightly its decision frontiers so that 
a set of specific queries convey the desired information. 
In the present paper, we formally introduce the problem and propose a
novel zero-bit watermarking algorithm that makes use
of \textit{adversarial model examples}.
While limiting the loss of performance of the protected model, this
algorithm allows subsequent extraction of the watermark using only few
queries. We experimented the approach on three neural networks designed for image classification, in the context of MNIST digit recognition task.

\end{abstract}

\section{Introduction}

Recent years have witnessed a fierce competition for the design and training of 
top notch deep neural networks.
%, with some notorious ones such as the ones by ~\cite{AAA,yt}. 
The industrial advantage from the possession
of a state of the art model is now widely acknowledged,
% (see \eg Amazon Machine Learning or BigML)
starting to motivate some attacks for stealing those models \cite{stealing,stealing2}.  Since it is now
clear that machine learning models will play a central
role in the IT development in the years to come, the necessity for
protecting those models appears more salient.

 In 1994, \cite{van1994digital} proposed to covertly embed a
  marker into digital content (such as audio or video data) in order
  to identify its ownership: by revealing the presence of such marker
  a copyright owner could prove its rights over the content. The
  authors coined the term digital watermarking. The fact that neural
  networks are digital content naturally questions the transferability
  of such techniques to those models.

\cite{Uchida:2017,Nagai2018} published the first method for watermarking a neural network
%certain type of neural networks (convolutional neural networks), 
that might be publicly shared and thus for which traceability through
ownership extraction is important. The marked object is here a
neural network and its trained parameters. However, this method
requires the ability to directly access the model weights: the model
is considered as a white box.
 The watermark embedding is
  performed through the use of a regularizer at training time. This regularization introduces the desired statistical bias into the parameters, which will serve as the watermark.
We are interested in a related though different problem, namely \textit{zero-bit
  watermarking} of neural networks (or any machine learning models) that are
only remotely accessible through an API.
The extraction of a zero-bit watermark in a given model refers to
  detecting the presence or the absence of the mark in that model. This type of watermark, along
with the required \emph{key} to extract it, is sufficient for an entity that
suspects a non legitimate usage of the marked model to confirm it or not.

In stark contrast to \cite{Uchida:2017,Nagai2018}'s approach, we seek a
black box watermarking approach that allows extraction to be conducted remotely,
without access to the model itself. 
More precisely, the extraction
test of the proposed watermark consists in a set of requests to the
machine learning service, available through an API \cite{stealing}. This allows the detection of (leaked) models
when model's parameters are directly accessible, but also
when the model is only exposed through an online service. 
Second, we
target the watermarking of models in general, \ie our scheme is not
restricted solely to neural networks, whether of a certain type or not.

\begin{figure*}[t!]
%  \vspace{-1cm}
  \center
  \includegraphics[width=.55\linewidth]{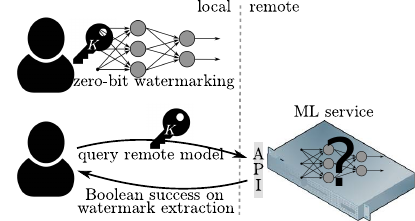}
%  \vspace{-1cm}
  \caption{Our goal: zero-bit watermarking a model locally (top-action), for
    remote assessment of a potential model leak (bottom-action).}
  \label{drawing}
\end{figure*}

\paragraph{Rationale.} We thus aim at embedding zero-bit watermarks into models, that can be extracted remotely.
In this setup, we can only rely on interactions with the model through
the remote API, \eg on object recognition queries in case of an image
classification model.  The input, \eg images, must thus convey a means
to embed identification information into the model (zero-bit
watermarking step) and to extract, or not, the identification
information from the remote model (watermark extraction step), see
Fig. \ref{drawing}.  Our algorithm's rationale is that the embedded
watermark is a slight modification of the original model's decision
frontiers around a set of specific inputs that form the
hidden \textit{key}.
%, thus operating in the model feature-space. 
Answers of the remote model to these inputs are compared to those of the marked model. 
A strong match (despite the possible manipulation of the leaked model) must indicate the presence of the watermark in the remote model with a high probability.

The inputs in the key must be crafted in a way that watermarking the model of interest
does not degrade significantly its performance. To this end,  
%In order to craft inputs that are not degrading significantly model performances, 
we leverage adversarial perturbations of training examples \cite{Goodfellow:2015}
that produce new examples (the ``adversaries'') very close the model's decision frontiers.
%that are generated on purpose to be close to model decision frontiers And 
As such adversaries tend to generalize across models, notably across different neural network architectures
for visual recognition, see \eg \cite{Rozsa:2016}, this frontier tweaking should resist model manipulation
and yield only few false positives (wrong identification of non marked models).

\paragraph{Contributions.} The contributions of this article are:
\textit{1)} A formalization of the problem of zero-bit watermarking a model
   for remote identification, and associated requirements (Section~\ref{sec:problem});
\textit{2)} A practical algorithm, the \textit{frontier stitching
     algorithm} based on adversaries that ``clamp'' the model frontiers, to address this problem. We also introduce a statistical framework for reasoning about the uncertainty regarding the remote model; we leverage a \textit{null hypothesis}, for measuring the success of the watermark extraction (Section~\ref{sec:algo});
\textit{3)} Experiments with three different types of neural networks on the MNIST dataset, validating
the approach with regards to the specified requirements (Section~\ref{sec:expes}).
%\end{itemize}

\section{Watermarking for Remote Extraction}
\label{sec:problem}

%\subsection{Considered scenario}
\paragraph{Considered scenario.}
The scenario that motivates our work is as follows: An entity, having
designed and trained a machine learning model, notably a neural
network, wants to zero-bit watermark it (top-action on
Fig.~\ref{drawing}). This model could then be placed in production
for applications and services. In case of the suspicion of a security breach in
that application (model has leaked by being copied at a bit-level), the entity suspecting a
given online service to re-use that leaked model can query that remote
service for answering its doubts (bottom-action).

Like for classic media watermarking methods (\cite{771066,van1994digital}), our 
approach includes operations of \textit{embedding} (the entity inserts the zero-bit watermark in its model),
and \textit{extraction} (the entity verifies the presence or not of
its watermark in the suspected model), and a study of possible \textit{attacks} (actions
performed by others in order to remove the watermark from the model).

\paragraph{Modeling Requirements.}

Following works in the multimedia domain \cite{771066}, and
by \cite{Uchida:2017,Nagai2018}, we adapt the requirements for a watermarking
method to the specific capability of \textit{remote} watermark
extraction (black box set-up). We choose those requirements to
structure the remaining of this article.

We consider the problem of zero-bit watermarking a generic classifier,
for remote watermark extraction. Let $d$ be the
dimension of the input space (raw signal space for neural nets or
hand-crafted feature space for linear and non-linear SVMs), and $C$
the finite set of target labels.  Let $k:\mathbb{R}^d\to C$ be
the \textit{perfect} classifier for the problem (\ie $k(x)$ is always
the correct answer).  Let $\hat{k}:\mathbb{R}^d\to C$ be the trained
classifier to be watermarked, and $F$ be the space of possible such
classifiers.
Our aim is to find a zero-bit watermarked version of
$\hat{k}$ (hereafter denoted $\hat{k}_w$) along with a set $K\subset \mathbb{R}^d$ 
of specific
inputs, named the \emph{key}, and their labels
$\{\hat{k}_w(x),x\in K\}$.
The purpose is to query with the key a remote model that can be either $\hat k_w$ or another unmarked model $k_r \in F$. 
The key, which is thus composed of ``objects'' to be classified, is used to embed the watermark into $\hat{k}$.

Here are listed the requirements of an \textit{ideal} watermarked model and key couple, $(\hat k_w,K)$:
\begin{description}
\item[Loyal.] The watermark embedding does not hinder the performance of the original
  classifier: 
  %$$\sum_{x\in X}||\hat{k(x)}-\hat{k}_w(x)||= 0.$$
  \begin{equation}
  \forall x \in \mathbb{R}^d, \notin K,~\hat{k}(x) = \hat{k}_w(x).
  \end{equation}
\item[Efficient.] The key is as short as possible, as accessing the watermark requires $|K|$ requests.
\item[Effective.] The embedding allows unique identification of 
  $\hat{k}_w$ using $K$ (zero-bit watermarking): 
  \begin{equation}
  \forall k_r\in F,~k_r\neq \hat{k}_w\Rightarrow
  \exists x\in K \text{ s.t. } k_r(x)\neq \hat{k}_w(x).
  \end{equation}
\item[Robust.] Attacks (such as fine-tuning or compression) to $\hat{k}_w$ do
  not remove the watermark\footnote{``$\hat{k}_w+\varepsilon$'' stands for a small modification of the parameters of $\hat{k}_w$ that preserves the value of the model, \ie that does not deteriorate significantly its performance.}:
  \begin{equation}
  \forall x\in K,~(\hat{k}_w+\varepsilon)(x) = \hat{k}_w(x).
  \end{equation}
\item[Secure.] No efficient algorithm exists to detect the
  presence of the watermark in a model by an unauthorized party.
\end{description}
Note that \textit{effectiveness} is a new requirement as compared to the list 
of Uchida \textit{et al}. Also, their  \textit{capacity} requirement, \ie the amount of
information that can be embedded by a method, is not part of ours
as our goal is to decide whether watermarked model is used or not (zero-bit watermark extraction).

One can observe the conflicting nature of %those requirements, specially 
effectiveness and robustness: If, for instance,
$(\hat{k}_w+\varepsilon)\in F$ then this function violates one of the two.
In order to allow for a practical setup for the problem, we rely on a
measure $m_K(a,b)$ of the matching between two classifiers $a,b\in F$:
\begin{equation}
  m_K(a,b)= \sum_{x\in K} \delta(a(x),b(x)),
  \label{eq:hamming}
\end{equation}
where $\delta$ is the
Kronecker delta. One can observe that $m_K(a,b)$ is simply the Hamming
distance between the vectors $a(K)$ and $b(K)$, thus based on elements
in $K$.
With this focus on distance, our two requirements can now be recast in a non-conflicting way:
\begin{itemize}
\item Robustness: $\forall \varepsilon \approx 0, m_K(\hat{k}_w,\hat{k}_w+\varepsilon) \approx 0$
\item Effectiveness: $\forall k_r\in F, m_K(\hat{k}_w,k_r) \approx |K|$
\end{itemize}

\section{The Frontier Stitching Algorithm}% for zero-bit watermarking}
\label{sec:algo}

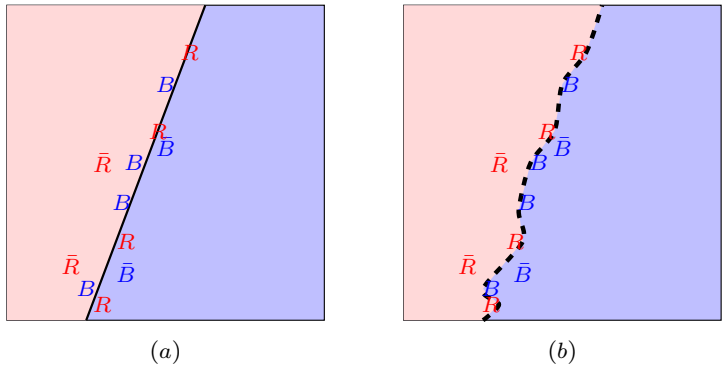
\begin{figure*}[t!]
  \center \usetikzlibrary{calc} \resizebox{0.55\textwidth}{!}{
  \begin{tikzpicture}[ every path/.style = {}, every node/.append style =
      {font=\sffamily} ]
      \begin{scope}[shift={(0,-5)}]

      \draw[fill=blue!25] (0,0) rectangle (4,4); \node at (2,-.4) {$(a)$};
      \fill[red!15] (0,0) -- (1,0) -- (2.5,4) -- (0,4) -- cycle;
      % \draw[ultra thick, dashed, smooth] (1,0)--(3.5,.1) -- (4,1) -- (0.1,3.1)
      % --
      % (0,3.6) -- (2.5,4);

      % \draw[ultra thick, dashed] plot[smooth] coordinates {(1,0) (1.2,.2) (1,.4)
      % (1.5,1) (1.45,1.4) (1.6,2) (2,3) (2.4,3) (2.3,3.4) (2.5,4) };

      \draw[thick] (1,0)--(2.5,4);

      % faux adversariaux
      \foreach \x/\y in {1.2/2,0.8/0.7} { \node[red] at (\x,\y)
        {$\bar R$}; }
      \foreach \x/\y in {1.5/0.6, 2/2.2} { \node[blue] at (\x,\y)
        {$\bar B$}; }

      %vrais adversariaux
      \foreach \x/\y in {1.2/0.2, 1.9/2.4, 2.3/3.4,1.5/1} { \node[red] at (\x,\y)
        {$R$}; }

      \foreach \x/\y in {1/0.4, 1.6/2, 2/3,1.45/1.5 } { \node[blue] at (\x,\y)
        {$B$}; }

      % \draw[thin, gray!50!red, dashed] (0,0)--(2.5,4);
      % \draw[thin, gray!50!red, dashed] (1,0)--(2,4);
      % \draw[thin, gray!50!red, dashed] (.4,0)--(4,4);
      % \draw[thin, gray!50!red, dashed] (0,0)--(2.5,4);
      % \draw[thin, gray!50!red, dashed] (0,1.6)--(4,2);
    \end{scope}

    \begin{scope}[shift={(5,-5)}]
      \draw[fill=blue!25] (0,0) rectangle (4,4); \node at (2,-.4) {$(b)$};
      % \fill[red!15] (0,0) -- (1,0) -- (2.5,4) -- (0,4) -- cycle;
      % \draw[ultra thick, dashed, smooth] (1,0)--(3.5,.1) -- (4,1) -- (0.1,3.1)
      % --
      % (0,3.6) -- (2.5,4);

      \fill[red!15] (0,0) --(1,0)-- (1.2,.2)-- (1,.4)-- (1.5,1)-- (1.45,1.4)--
      (1.6,2)-- (1.9,2.4)-- (2,3)-- (2.3,3.4) --(2.5,4) --(0,4) --cycle;

      % \draw[thick, gray!50!green, dashed] (1,0)--(2.5,4);
      % \draw[thin, gray!50!red, dashed] (1,0)--(2,4);
 %     \draw[thin, gray!50!red, dashed] (.4,0)--(4,4); \draw[thin, gray!50!red,
 %       dashed] (0,1.6)--(4,2);

      \draw[ ultra thick, dashed] plot[smooth] coordinates {(1,0) (1.2,.2) (1,.4) (1.5,1)
        (1.45,1.4) (1.6,2) (1.9,2.4) (2,3) (2.3,3.4) (2.5,4) };

     % \draw[ultra thick] (1,0)--(2.5,4);
      \foreach \x/\y in {1.2/2,0.8/0.7} { \node[red] at (\x,\y)
        {$\bar R$}; }
      \foreach \x/\y in {1.5/0.6, 2/2.2} { \node[blue] at (\x,\y)
        {$\bar B$}; }

      \foreach \x/\y in {1.2/0.2, 1.9/2.4, 2.3/3.4,1.5/1} { \node[red] at
        ($(\x,\y)+(-.1,0)$) {$R$}; }

      \foreach \x/\y in {1/0.4, 1.6/2, 2/3,1.45/1.5 } { \node[blue] at
        ($(\x,\y)+(.1,0)$) {$B$}; }

      % \draw[thin, gray!50!red, dashed] (0,0)--(2.5,4);
      % \draw[thin, gray!50!red, dashed] (0,1.6)--(4,2);
    \end{scope}
      \end{tikzpicture}
  }%end resizebox
%\vspace{-.3cm}
\caption{Illustration of proposed approach for a binary
    classifier.
    (a) The proposed algorithm
    first computes ``true adversaries'' ($R$ and $B$) and 
    ``false'' ones ($\bar{R}$ and $\bar{B}$) for both classes from training examples.
    They all lie close the decision frontier.  
    (b) It then fine-tunes the classifier such that these inputs are now all well classified, 
    \ie\ the 8 true adversaries are now correctly classified in this example while the 4 false ones remain so. This 
    results in a loyal watermarked model (very similar to original one) and a key size of
    $|K|=12$ here.
    This process resembles ``stitching'' around
    data points.} %inspiring the name of our proposed algorithm.}
\label{illustration}
\end{figure*}

We now present a practical zero-bit model watermarking algorithm that
permits remote extraction through requests to an API, following the
previously introduced requirements.  Our aim is to output a
watermarked model $\hat{k}_w$, which can for instance be placed into
production for use by consumers, together with a watermark key $K$ to
be used in case of model leak suspicion. For the security
  requirement to hold, we obviously discard any form of \textit{visible}
  watermark insertion \cite{visible}.  Fig.~\ref{illustration}
illustrates the approach in the setting of a binary classifier
(without loss of generality).

As we use input points for watermarking the owned model and subsequently to query a suspected remote model, the
choice of those inputs is crucial. 
A non watermarking-based solution based simply on choosing arbitrarily $|K|$ training examples (along with their correct labels),
is very unlikely to succeed in the identification of a specific valuable model:
Classifying those points correctly should be easy for highly accurate
classifiers, which will then provide similar results, ruining the effectiveness.
On the other hand, the opposite strategy of selecting $|K|$
arbitrary examples and fine-tuning $\hat{k}$ so that it changes the way
they are classified
%those points are evaluated to a different class than the one expected
(\eg\ $\forall x\in K,~\hat{k}(x)\neq \hat{k}_w(x)$) 
is an option to modify the model's behavior in an identifiable way. 
%frontier tweak that may identity the watermarked network. 
However, fine-tuning on even few examples that are possibly far from decision frontiers
%those extreme datapoints (therefore including a watermark in that model) 
will significantly alter the performance 
%bend the decision frontiers 
of $\hat{k}$: The produced solution will not be loyal.

Together, those observations lead to the conclusion that the selected
points should be close to the original model's decision frontier, that is,
their classification is not trivial and depends heavily on the model (Fig.~\ref{illustration}(a)). 
Finding and manipulating such inputs is the purpose of adversarial perturbations \cite{Goodfellow:2015,Moosavi-Dezfooli:2017}. 
Given a trained model, any well classified example can be modified in a very slight way such that it is now
misclassified.
%For instance, a natural image that is correctly recognized by a given model can be modified 
%in an imperceptible way so as to be assigned a wrong class.
Such modified samples are called ``adversarial examples'', or adversaries in short.

The proposed frontier stitching algorithm, presented in
Algorithm~\ref{algo-marking}, makes use of such adversaries,
selected to ``clamp'' the frontier in a unique, yet harmless way (Fig.~\ref{illustration}(a)). It proceeds 
in two steps to mark the model.
%We proceed in the following
%way. For the watermark embedding, we will pursue a two-step approach. 
The first step is to select a small key set $K$ of specific input
points, which is composed of two types of adversaries.
It first contains classic adversaries, we call \textit{true adversaries}, that are misclassified by $\hat{k}$ although being each very close to a well classified example. It also contains \textit{false adversaries}, each obtained by applying an adversarial perturbation to a well classified example without ruining its classification. In practice, the ``fast gradient sign method'' proposed in \cite{Goodfellow:2015} is used with a suitable gradient step to create potential adversaries of both types from training examples, these adversaries are inputs that will be closer to a decision frontier than their base inputs. This modification in the direction of other classes is the purpose of adversarial attacks. (Other methods include for instance the ``Jacobian-based saliency map approach'' \cite{jsma}).

These frontier clamping inputs are then used to mark the model (Fig.~\ref{illustration}(b)). The model $\hat{k}$ is fine-tuned into $\hat{k}_w$ such that all points in $K$ are now well classified: 
\begin{equation}
\forall x \in K,~\hat{k}_w(x)=k(x).
\end{equation} 
In other words, the true adversaries of $\hat{k}$ in $K$ become false adversaries of the marked model, and false adversaries remain as such. The role of the false adversaries is to limit strongly the amount of changes that the decision frontiers will undergo when getting true adversaries back to the right classes. False adversaries also have the role of characterizing the shapes of the model frontiers, for adding robustness to the statistical watermark extraction process we now present.
%using $K$ and the desired labels $K_{labels} = \{\hat{k}_w(x),x\in K\}$ as input to
%the training, as presented on Figure~\ref{illustration}(d).

\paragraph{Statistical watermark extraction.}
The marking step is thus the embedding of such a crafted key in
the original model, while the watermark extraction consists in asking
the remote model to classify the inputs in key $K$, to assess the
presence or not of the zero-bit watermark
(Algorithm~\ref{algo-querying}). We now analyze statistically this
extraction problem.

As discussed in Section \ref{sec:problem}, the key quantity at extraction time
is the Hamming distance $m_K$ (Eq. \ref{eq:hamming}) between remote model's
answers to the key and expected answers.  The stitching algorithm produces
deterministic results with respect to the imprinting of the key: A marked model
perfectly matches the key, \ie the distance $m_k$ between $\hat k_w$ and query results in $\text{Algorithm~2}(K,K_{labels})$ equals zero.   
%$m_K(\hat k_w,\text{Algorithm~2}(K,K_{labels}))=0$.
However, as the leaked model
may undergo arbitrary attacks (\eg for watermark removal), transforming
$\hat k_w$ into a model $\hat k_w'$, one should expect some deviation in
$\hat k_w'$ answers to watermark extraction
($0\leqslant m_K(\hat k_w,\hat k_w')\ll |K|$). On the other hand, other unmarked
models might also partly match key labels, and thus have a positive non-maximum
distance too.  As an extreme example, even a strawman model that answers a label
uniformly at random produces $\vert K \vert/|C|$ matches in expectation, when
classifying over $|C|$ classes. Consequently, two questions are central to the
frontier stitching approach: \emph{How large is the deviation one should tolerate from
the original watermark in order to state about successful zero-bit watermark?}
And, dependently, \emph{how large should the key be, so that the tolerance is
increased?}

\begin{algorithm*}[t!]
\normalsize
\caption{Zero-bit watermarking a model} %, with frontier stitching}
  \label{algo-marking}
 \begin{algorithmic}[1]
   \REQUIRE Labelled sample set $(X,Y)$; Trained model $\hat k$; Key length $\ell = |K|$; Step size $\varepsilon$ for adversary generation; 
   %data tensor $V_X$ and corresponding labels $V_Y$.
   \ENSURE $\hat k_w$ is watermarked with key $K$
   %   \STATE $key_{true} \gets$ generate $l \times r$ adversarial examples using $\varepsilon$ step on $V_X$ (cf. \cite{Goodfellow:2015})
   \COMMENT{Assumes $X$ is large enough and $\varepsilon$ is balanced to generate true \& false adversaries}

   \COMMENT{Key construction}
   %   \STATE $adv\_candidates \gets$ \textsc{FAST\_GRADIENT\_SIGN\_METHOD($\hat k$,$V_X$,$\varepsilon$)} {~~~\scriptsize \cite{Goodfellow:2015}}
      \STATE $adv\_candidates \gets$ \textsc{GEN\_ADVERSARIES($\hat k$,$(X,Y)$,$\varepsilon$)} 
   \WHILE{$|key_{true}|<\ell/2 $ or $ |key_{false}|<\ell/2$}
    \STATE {pick random adversary candidate $c\in adv\_candidates$, associated to $x\in X$ with label $y_x$}

 %   \COMMENT{$c$ is indeed an adversarial example}
    \IF[$c$ is a true adversary]{$\hat{k}(x) = y_x$ and $\hat{k}(c) \neq y_x$ and $|key_{true}|<\ell/2 $}
    \STATE {$key_{true} \gets key_{true}\cup \{(c,y_x)\} $}
    \ELSIF[$c$ is a false adversary]{ $\hat{k}(c)= \hat{k}(x) = y_x$ and $ |key_{false}|<\ell/2$ }
%    \COMMENT{$c$ is not an adversarial example}
    \STATE {$key_{false}\gets key_{false} \cup \{(c,y_x)\}$}
    \ENDIF
   \ENDWHILE
   \STATE $(K,K_{labels}) \gets key_{true} \cup key_{false}$
   
%   \COMMENT{Key embedding}
%   \STATE $K_{labels} \gets \{ k(c), \forall c \in K \}$
%
   \COMMENT{Force embedding of key adversaries in their original class}
   \STATE $\hat k_w \gets$ \textsc{TRAIN($\hat k$, $K$, $K_{labels}$)} 
   \RETURN $\hat k_w$, $K$, $K_{labels}$
  \end{algorithmic}
\end{algorithm*}

\begin{algorithm*}[t!]
\normalsize
  \caption{Zero-bit watermark extraction from a remote model}
  \label{algo-querying}
 \begin{algorithmic}[1]
   \REQUIRE $K$ and $K_{labels}$, the key and labels used to watermark the neural network
   \STATE{ $m_K \gets 0$}
   \FOR{\textbf{each} $c \in K$}
   \IF{\textsc{QUERY\_REMOTE($c$) $\neq K_{labels}(c)$}}
   \STATE {$m_K \gets m_K+1$ } \COMMENT{remote model answer differs from
     recorded answer}
   \ENDIF
   \ENDFOR
%   \RETURN $d \gets$ \textsc{HAMMING-DISTANCE($K_{labels}$,$vector$)}
   %   \COMMENT{the lower $d$ is, the most probable the remote neural network is the watermarked one}
   
\COMMENT{Having $\theta $ such that $ 2^{-\vert K\vert}\sum_{z=0}^\theta{\vert K\vert\choose z}<0.05$ ~(\textit{null-model})}

   \RETURN \textsc{$m_K< \theta$}\COMMENT{True $\Leftrightarrow$ successful extraction}
  \end{algorithmic}
\end{algorithm*}

%What is then the criteria to decide whether a remote model is a
%variation of our marked model or not ?
%Since even the strawman model has a small chance of producing exactly
%the key, one can only
We propose to rely on a probabilistic approach by estimating the probability of
an unmarked model $k_r$ to produce correct answers to requests from inputs in the
key.
%, \ie to have $m_K(\hat k_w,r)>0$. 
While providing an analysis that would both be precise and cover all model
behaviors is unrealistic, we rely on a \emph{null-model} assuming that inputs in
the key are so close to the frontier that, at this ``resolution'', the frontier
only delimits two classes (the other classes being too far from the considered
key inputs), and that the probability of each of the two classes is $1/2$
each. This is all the more plausible since we leverage adversaries especially
designed to cause misclassification.

%, while those inputs are by the two class frontiers (the
%original class and the class causing misclassification), we examine
%this hypothesis as plausible in the rest of this paper.
%\gilles{strengthen motivation for null-model.}

More formally, let $k_\emptyset$ be the null-model:
$\forall x\in K,~ \mathbb{P}[k_\emptyset(x)=\hat{k}_w(x)]=1/2$.
%\gilles{some may say it's simple but it's
%  the only one we've got}.
Having such a null-model allows applying a $p$-value approach to the
decision criteria. Indeed, let $Z=m_K(\hat k_w,k_r)$ be the random variable
representing the number of mismatching labels for key inputs $K$.
%$$ X = m_K(a,b)=  \sum_{x\in K} \delta(K(x),r(x)),$$
%$$X= \sum_{x\in K} \delta(k(x),r(x))$$.
Assuming that the remote model is the
null-model, the probability of having exactly $z$ errors in the key is
$\mathbb{P}[Z=z|k_r=k_\emptyset] = 2^{-\vert K\vert}{\vert K\vert \choose z}$, that is $Z$ follows the binomial
distribution $B(|K|,\frac{1}{2})$. %of size $\vert K\vert$.
Let $\theta$ be the maximum number of errors tolerated on $k_r$'s
answers to decide whether or not the watermark extraction is successful. To
safely ($p$-value $<0.05$) reject the hypothesis that $k_r$ is a model
behaving like our null-model, we need $\mathbb{P}[Z\leq\theta|k_r=k_\emptyset]<0.05$. That is $2^{-\vert K\vert}\sum_{z=0}^\theta{\vert
  K\vert\choose z}<0.05$. In particular, for  key sizes of $|K|=100$ and $|K|=20$
a $p$-value of $0.05$, the maximum number of tolerated errors are
$\theta=42$ and $6$, respectively.
%\gilles{Although it is pretty arbitrary, the experimental
%  section will show it works well in practice.}
We thus consider the zero-bit watermark extraction from the remote
model successful if the number of errors is below that
threshold $\theta$, as presented in
Algorithm~\ref{algo-querying}. Next Section includes an experimental study of
false positives related to this probabilistic approach.

\section{Experiments}
\label{sec:expes}

We now conduct experiments to evaluate the proposed approach in the
light of the requirements stated in Section~\ref{sec:problem}.  In
particular, we evaluate the fidelity, the effectiveness and the
robustness of our algorithm.

%\subsection{Evaluation Settings}

%% \begin{table}
%%   \centering
%%   \begin{tabular}{|l|l|l|l|}
%%     \hline
%%     ~ & \#Parameters  & Details & Accuracy \\ \hline
%%     CNN & $710,218$  & mnist\_cnn.py & 0.993 (10 epochs) \\
%%     IRNN & $199,434$ & mnist\_irnn.py & 0.9918 (900 epochs) \\
%%     MLP & $669,706$ & mnist\_mlp.py & 0.984 (10 epochs) \\
%%     \hline
%%   \end{tabular}
%%   \caption{Neural networks used for experiments on the MNIST dataset.}
%%   \label{NNs}
%%   \end{table}

%\paragraph{MNIST classifiers.}

We perform our experiments on the MNIST dataset \cite{lecun1998mnist}, using the Keras
backend \cite{keras} %\footnote{\url{https://keras.io/backend/}}
to the TensorFlow platform\footnote{Code will be open-sourced on GitHub, upon article acceptance.} \cite{tensorflow2015-whitepaper}.
%for running our different experiments.
As neural network architectures, we use three off-the-shelf
implementations, available publicly on the Keras website, namely
  {\tt mnist\_mlp} (0.984\% accuracy at 10 epochs, we denote as MLP),
  {\tt mnist\_cnn} (0.993\% at 10, denoted as CNN) and {\tt
    mnist\_irnn} (0.9918\% at 900, denoted as IRNN).  Their
  characteristics are as follows. The MLP is composed of two fully
  connected hidden layers of 512 neurons each, for a total of
  $669,706$ parameters to train. The CNN is composed by two
  convolutional layers (of size 32 and 64), with kernel sizes of
  $3\times3$, followed by a fully connected layer of 128 neurons (for a
  total of $710,218$ parameters). Finally, the IRNN refer to settings
  by Le \textit{et al}. \cite{DBLP:journals/corr/LeJH15} and uses a fully
  connected recurrent layer (for a total of $199,434$ parameters).
  All three architectures use a softmax layer as output.

All experiments are run on networks
trained with the standard parametrization setup: MNIST training set of
$60,000$ images, test set of size $10,000$, SGD with mini-batches of size $128$ and a learning rate of
$0.001$.
\subsection{Generating adversaries for the watermark key.}
We use the Cleverhans Python library
by \cite{papernot2017cleverhans} % \footnote{\url{https://github.com/tensorflow/cleverhans}}
to generate the adversaries (function \textsc{GEN\_ADVERSARIES()} in
Algorithm~\ref{algo-marking}). It implements the ``fast gradient sign
method'' by \cite{Goodfellow:2015}, we recall here for completeness.
With $\theta$ the parameters of the attacked model, $J(\theta,x,y)$ the cost function used to train the model and $\bigtriangledown$ the gradient of that cost function with respect to input $x$, the adversarial image $x^*$ is obtained from the input image $x$ by applying the following perturbation:
$$x^*=x+\varepsilon~.~sign(\bigtriangledown_{x} J(\theta,x,y)).$$
  $\varepsilon$ thus controls the intensity of the adversarial perturbation; we use a default setting of $\varepsilon=0.25$.
  
Alternative methods, such as the
``Jacobian-based saliency map'' (\cite{2015arXiv151107528P}), or other attacks for generating adversaries may also be used (please refer to \cite{adv-survey} for a list of existing approaches).
  Adversaries are crafted
from some test set images that are then removed from the test set to avoid biased results.

\begin{figure}
  \centering
%  \fcolorbox{red}{red}{
    \subfigure[]{\includegraphics[width=0.24\linewidth]{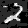}}
    \subfigure[]{\includegraphics[width=0.24\linewidth]{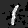}} 
    \subfigure[]{\includegraphics[width=0.24\linewidth]{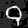}}
    \subfigure[]{\includegraphics[width=0.24\linewidth]{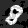}}
%}
    \caption{An example of a key $K$ of size four (generated with the ``fast gradient sign method'' and $\varepsilon=0.1$), for the CNN classifier and the MNIST task: Key inputs (a) and (b) are true adversaries (2 and 1 digits, classified as 9 and 8 respectively), while (c) and (d) are false adversaries (9 and 8 digits that are correctly classified but knowingly close to a decision frontier).}
    \label{fig:key}
\end{figure}

As explained in Section \ref{sec:algo}, our key set $K$ is composed by
  $50\%$ of true adversaries, and by $50\%$ of false adversaries
  (adversarial perturbations that are not causing misclassification).
In the fast gradient sign method, the value of $\varepsilon$ controls the intensity of the perturbations that are applied to the attacked images. With a large value, most of the adversaries are true adversaries and, conversely, a small $\varepsilon$ produces mostly false adversaries. As a consequence, $\varepsilon$ must be chosen so that Algorithm~\ref{algo-marking} as enough inputs ($\ell/2$) of each kind, in order to build the key.
Altogether, the adversaries in the key are close to the decision frontiers, so that they ``clamp'' these boundaries. An example of a key is displayed in Fig. \ref{fig:key}.

\subsection{Impact of watermarking (fidelity requirement).}
This experiment considers the impact on fidelity of the watermark embedding, of
sizes $|K|=20$ and $|K|=100$, in the three networks. We generated multiple keys
for this experiment and the following ones (see Algorithm~\ref{algo-marking}),
and kept those which required fewer that $100$ epochs for embedding in the
models (resp. $1000$ for IRNN), using a fine tuning rate of ${1}\over{10}$th of
the original training rate.  The following results are
averaged over $30$ independent markings per network.

\begin{figure*}[]
  \center
%    \fcolorbox{red}{red}{
      \includegraphics[width=0.75\textwidth]{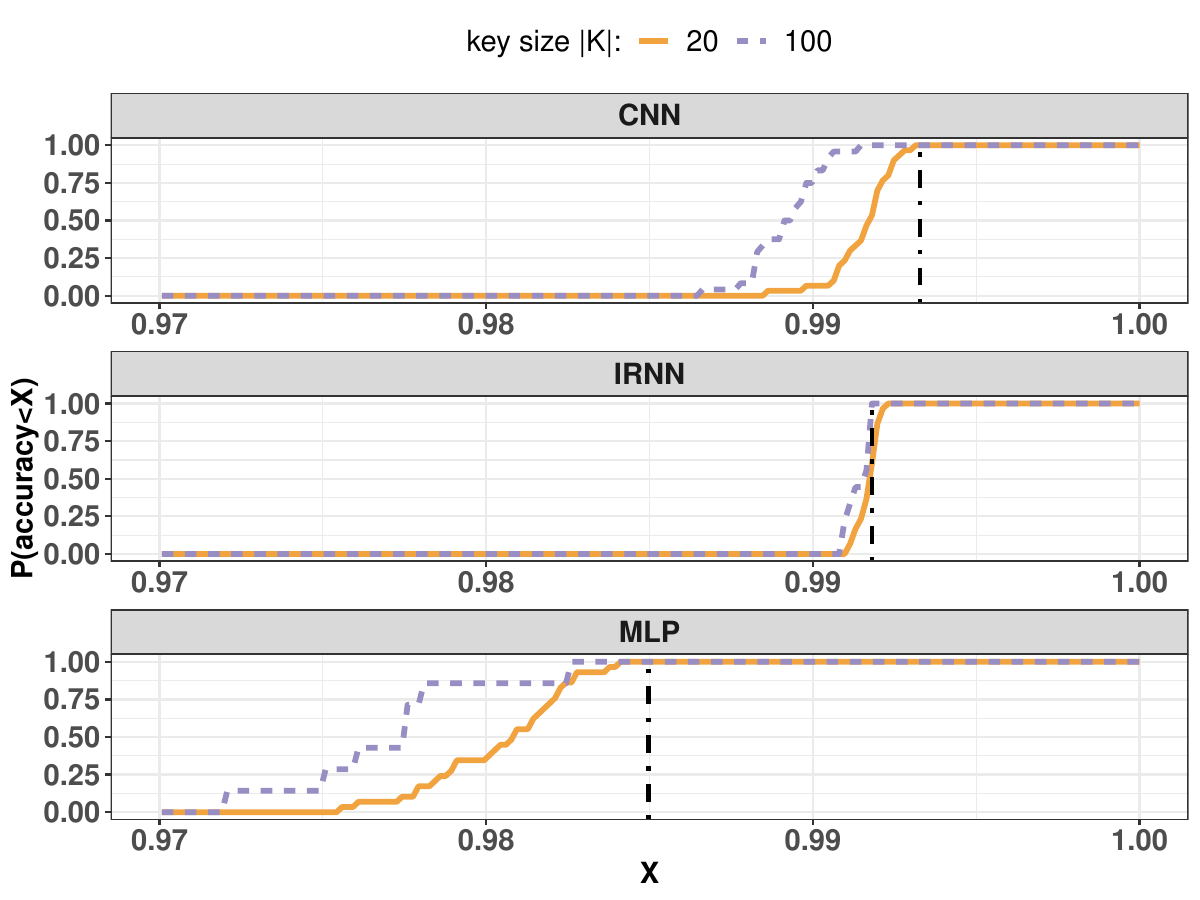}
 %     }
%\vspace{-.3cm}
\caption{Distribution of the degradation caused by watermarking models (resulting accuracy), with %$|K| \in \{20,100\}$%
multiple keys of size $20$ and $100$ ($\varepsilon=0.25$).
  %Pre-watermarking accuracies are indicated with v5$tical dot-dash lines.
   Black vertical dot-dash lines indicate pre-watermarking accuracies.
  }
    \label{fidelity}
\end{figure*}

%% The first question is wether it is possible to watermark a traning
%% model with adversarial examples in the watermark key, without
%% sigificant accuracy degration that would make the model useless.  This
%% question is far from trivial, since it is for example shown in
%% \cite{Uchida:2017} that direct modifications (\ie without training) of
%% parameters of the learned model cause watermark embedding failure.
The cumulative distribution function (CDF) in Fig. \ref{fidelity} shows
the accuracy for the 3 networks after embedding keys of the two sizes. 
%watermark embedding
%corresponding to those two key size, and for the three networks.  
IRNN exhibits nearly no degradation, while embedding in the MLP causes
on average $0.4\%$ and $0.8\%$ loss for respectively key sizes $20$ and
$100$.

We remarked no significant degradation difference when
  marking ($|K|=20$, 10 independent runs) a model with adversaries generated under $\ell_1$, $\ell_2$ and $\ell_{\infty}$ norms. For instance, we marked the CNN model with $\varepsilon=150$ ($\ell_1$, fooling $88.54\%$ of MNIST test set), $\varepsilon=7$ ($\ell_2$, fooling $87.56\%$) and $\varepsilon=0.25$ ($\ell_\infty$, fooling $89.08\%$); This has resulted in accuracy drops of respectively $0.23\%$,  $0.22\%$ and $0.14\%$. We use $\ell_{\infty}$ in the sequel.

%  keyS   ml       moy           sd
% 1   20  CNN 0.9915185 0.0009232930
% 2   20 IRNN 0.9916768 0.0003122931
% 3   20  MLP 0.9803622 0.0021834338
% 4  100  CNN 0.9891625 0.0010925828
% 5  100 IRNN 0.9913805 0.0003745555
% 6  100  MLP 0.9769553 0.0032467212

%\subsection{False positives in remote watermark extraction (effectiveness requirement)}
\subsection{False positives in remote watermark extraction (effectiveness requirement).}
\label{ss:fp}
\begin{figure*}[h!]
\center
\includegraphics[width=0.8\textwidth]{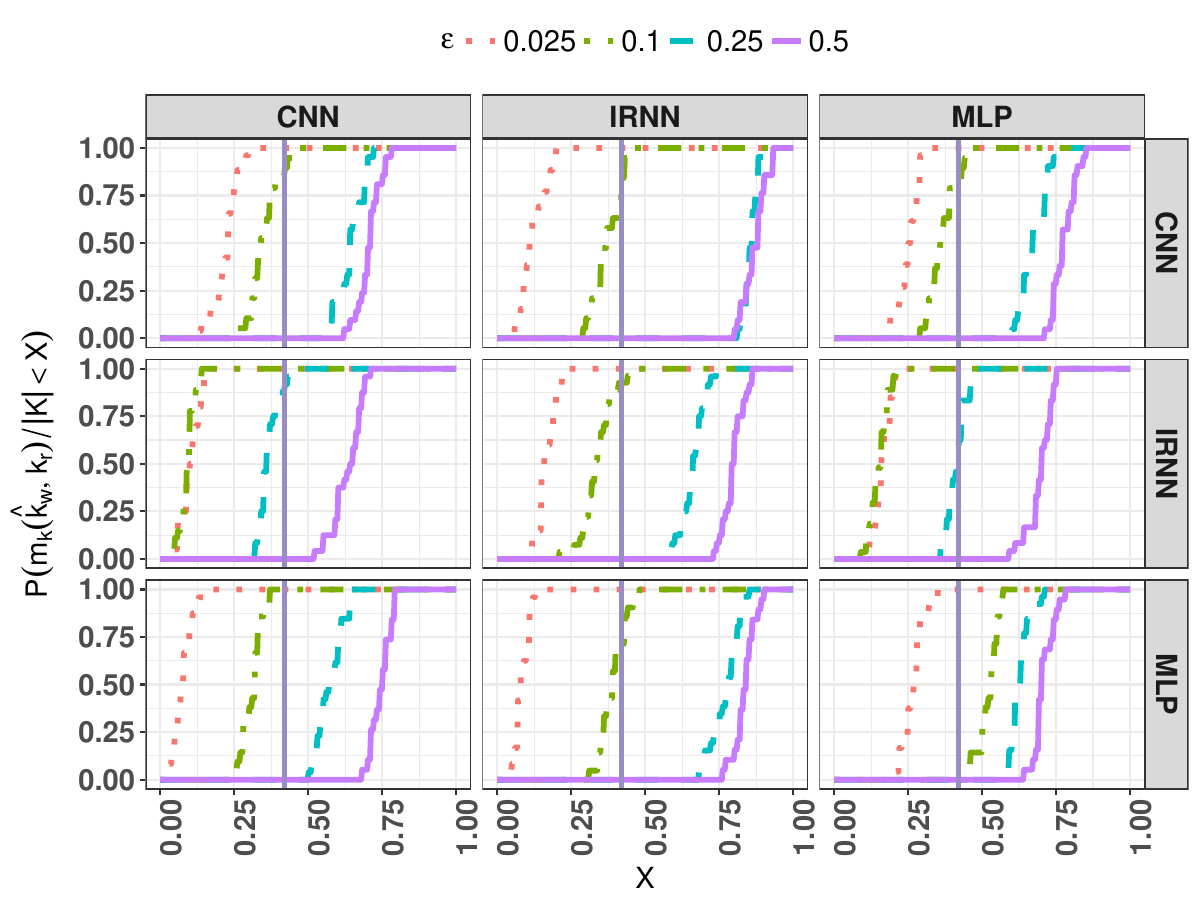}
\caption{
    Distribution of (normalized) Hamming distance when
      extracting the watermark (generated on networks $\hat k_w$ listed on right axis) from
      remote unmarked models $k_r$ (w. names on the top), for $|K|=100$
      and four values of $\varepsilon$. Vertical solid bars indicate the decision threshold
      corresponding to $p<.05$ criteria. False positives
      (\ie\ unmarked model at distance less than the
      threshold) happen when the CDF is on the left of the bar.
      Algorithm~\ref{algo-marking}'s value $\varepsilon=0.5$
      (purple solid curve) prevents the issue of false positives in all cases.}
    \label{epsilon}
\end{figure*}

 %We now measure Algorithm~\ref{algo-querying} false positive rate.
  We now experiment the effectiveness of the watermark extraction (Algorithm~\ref{algo-querying}).
  When querying the remote model returns True, it is important to get
  a low false positive rate.  To measure this, we ran on \textit{non
  watermarked} and retrained
  networks of each type
the extraction Algorithm~\ref{algo-querying}, with keys used to watermark
the three original networks. 
Ideally, the output should always be negative. 
We use $|K|=100$, and various values of
$\varepsilon \in \{0.025,0.1,0.25,0.5\}$.
We observe on Fig.~\ref{epsilon} that the
false positives are occurring for lower values
$0.025$ and $0.1$ on some scenarios. False positives disappear for
$\varepsilon=0.5$.

\begin{figure}[h!]
    \centering
%    \fcolorbox{red}{red}{
      \includegraphics[width=0.85\linewidth]{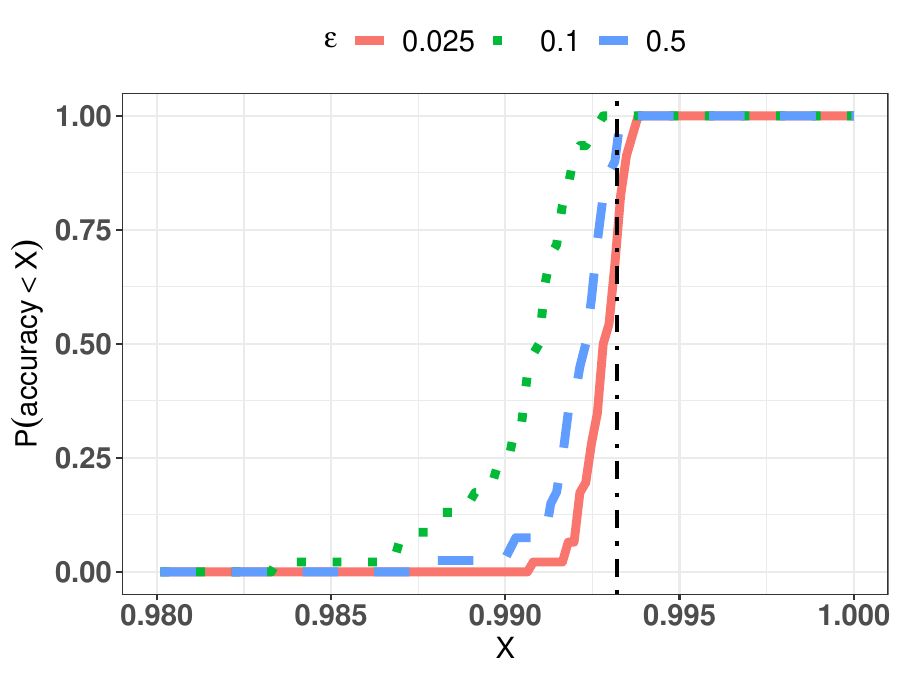}
 %    }
   \caption{Distribution of the degradation caused by watermarking the CNN model (resulting accuracy), with multiple keys of size 20, and for various $\varepsilon \in
     \{0.025,0.1,0.5\}$ in the $\ell_\infty$ norm.   The black vertical dot-dash line indicates the pre-watermarking accuracy.}
    \label{fig:marking-epsilon}
\end{figure}

 This last experiment indicates that the model owner has to
  select a high $\varepsilon$ value, depending on her dataset, as the
  generated adversaries are powerful enough to prevent accurate
  classification by the remote inspected model. We could not assess a
  significant trend for a higher degradation of the marked model when
  using a higher $\varepsilon$ as depicted in
  Fig. \ref{fig:marking-epsilon}, where the CNN model is marked with
  keys of $|K|=20$ and for $\varepsilon \in \{0.025,0.1,0.5\}$ (10
  runs per $\varepsilon$ value). The $0.25$ value is to be observed on
  Fig. \ref{fidelity}.  We note that this relates to
  \textit{adversarial training}, a form of specialized data
  augmentation that can be used as generic regularization
  \cite{Goodfellow:2015} or to improve model resilience to adversarial
  attacks \cite{papernot2017cleverhans}. Models are thus trained with
  adversarial examples, which is in relation to our watermarking
  technique that incorporates adversarial examples in a fine-tuning
  step, without ruining model accuracy.

\subsection{Attacking the watermarks of a leaked model (robustness requirement).}

We now address the robustness of the stitching algorithm.  Two types
of attacks are presented: Model compression (via both \textit{pruning}
and \textit{singular value decomposition}) and overwriting via
fine-tuning.

We consider \textit{plausible} attacks over the leaked model,
\ie\ attacks that do not degrade the model beyond a certain accuracy,
which we set to $0.95$ in the sequel\footnote{This about $3.5\%$
  accuracy drop is also the one tolerated by a recent work on
  \textit{trojaning} neural networks~\cite{trojaning}.} (the three networks in our
experiments have accuracy above $0.984$).

In our set-up, re-using a leaked model that has been
  significantly degraded in the hope to remove a possible watermark
  does not make sense; the attacker would rather use a less
  precise, yet legitimate model.

We remark that due to the nature of our watermarking method, an
attacker (who does not possess the watermark key) will not know whether
or not her attacks removed the watermark from the leaked model. 

\textit{Compression attack via pruning}\quad As done by \cite{Uchida:2017}, we study the effect of compression
through parameter pruning, where $25\%$ to $85\%$ of model weights
with lowest absolute values are set to zero.  Results are presented on
Tab.~\ref{compression}. Among all plausible attacks, none but one
($50\%$ pruning of IRNN parameters) prevents successful and $100\%$
accurate extraction of the watermarks.
We note that the MLP is prone
to important degradation of accuracy when pruned, while at the
same time the average number of erased key elements from the model is
way below the decision threshold of $42$. Regarding the CNN, even
$85\%$ of pruned parameters are not enough to reach that same
threshold.

%%%%%%% FULL TABLE WITH STDEV ACC column
%% \newcolumntype{g}{>{\columncolor{green!10}}c}
%% \begin{table}
%% \center
%% \resizebox{0.53\textwidth}{!}{
%%    \begin{tabular}{|c|cccgll|}
%%     \hline
%%     ~ & Pruning rate & Avg $K$ elts rem. & Stdev & \cellcolor{white!0} Extraction rate & Acc. after & Acc. stdev \\ \hline
%%    CNN  &    0.25 &   0.053/100 &  0.229 &  1.000 &  0.983 &  0.003 \\
%%      -  &    0.50 &   0.263/100 &  0.562 &  1.000 &  0.984 &  0.002 \\
%%      -  &    0.75 &   3.579/100 &  2.479 &  1.000 &  0.983 &  0.003 \\  \rowcolor{Gray}
%%      -  &    0.85 &  34.000/100 &  9.298 &  0.789 &  0.936 &  0.030 \\
%%  IRNN  &    0.25 &  14.038/100 &  3.873 &  1.000 &  0.991 &  0.001 \\
%%      -  &    0.50 &  59.920/100 &  6.782 &\cellcolor{red!10}  0.000 &  0.987 &  0.001 \\ \rowcolor{Gray}
%%      -  &    0.75 &  84.400/100 &  4.093 &  0.000 &  0.148 &  0.021 \\
%%    MLP  &    0.25 &   0.360/100 &  0.700 &  1.000 &  0.951 &  0.018 \\  \rowcolor{Gray}
%%      -  &    0.50 &   0.704/100 &  0.724 &  1.000 &  0.947 &  0.021 \\  \rowcolor{Gray}
%%      -  &    0.75 &   9.615/100 &  4.392 &  1.000 &  0.915 &  0.031 \\  \rowcolor{Gray}
%%      -  &    0.80 &  24.438/100 &  5.501 &  1.000 &  0.877 &  0.042 \\
%%     \hline
%% \end{tabular}
%% }
\newcolumntype{g}{>{\columncolor{green!10}}c}
\begin{table*}[h]
\center
\resizebox{0.6\textwidth}{!}{
   \hspace{-0.3cm}\begin{tabular}{|c|ccccl|}
    \hline
    ~ & Pruning rate & $K$ elts removed & Stdev & \cellcolor{white!0} Extraction rate & Acc. after\\ \hline
   CNN  &    0.25 &   0.053/100 &  0.229 &  \cellcolor{green!10} 1.000 &  0.983 \\
     -  &    0.50 &   0.263/100 &  0.562 &  \cellcolor{green!10} 1.000 &  0.984 \\
     -  &    0.75 &   3.579/100 &  2.479 &  \cellcolor{green!10} 1.000 &  0.983 \\  \rowcolor{Gray}
     -  &    0.85 &  34.000/100 &  9.298 &  0.789 &  0.936 \\
 IRNN  &    0.25 &  14.038/100 &  3.873 &  \cellcolor{green!10} 1.000 &  0.991  \\
     -  &    0.50 &  59.920/100 &  6.782 &\cellcolor{red!10}  0.000 &  0.987  \\ \rowcolor{Gray}
     -  &    0.75 &  84.400/100 &  4.093 &  0.000 &  0.148 \\
   MLP  &    0.25 &   0.360/100 &  0.700 &  \cellcolor{green!10} 1.000 &  0.951 \\  \rowcolor{Gray}
     -  &    0.50 &   0.704/100 &  0.724 &  1.000 &  0.947 \\  \rowcolor{Gray}
     -  &    0.75 &   9.615/100 &  4.392 &  1.000 &  0.915 \\  \rowcolor{Gray}
     -  &    0.80 &  24.438/100 &  5.501 &  1.000 &  0.877 \\
    \hline
\end{tabular}
}
%\vspace{-0.2cm}
\caption{Robustness to compression attack (pruning-based): Watermark extraction
  success rates ($|K|=100$), after a pruning attack on watermarked
  models. Different pruning rates are tested to check if watermarks get erased
  while accuracy remains acceptable. Results in gray rows are to be ignored (not plausible attacks).}
%  degrading the model accuracy beyond admissible threshold $\delta$.}
  \label{compression}
\end{table*}

\textit{Compression attack via Singular Value Decomposition}\quad

\newcolumntype{g}{>{\columncolor{green!10}}c}
\begin{table*}
\center
\resizebox{0.6\textwidth}{!}{
   \hspace{-0.3cm}\begin{tabular}{|c|ccccl|}
    \hline
    ~ & Pruning rate & $K$ elts removed & Stdev & \cellcolor{white!0} Extraction rate & Acc. after\\ \hline
   CNN  &    0.25 &   1.867/100  & 1.457 &  \cellcolor{green!10} 1.000 &  0.987 \\  \rowcolor{Gray}
     -  &    0.50 &   20.143/100 & 4.721 &  1.000 &  0.885 \\ \rowcolor{Gray}
     -  &    0.75 &   69.643/100 & 3.296 &  0.000 &  0.426 \\ \rowcolor{Gray}
     -  &    0.90 &   83.714/100 & 3.989 &  0.000 &  0.255 \\
   MLP  &    0.25 &   0.385/100  & 0.898 &  \cellcolor{green!10} 1.000 &  0.973 \\ 
     -  &    0.50 &   5.760/100  & 2.697 &  \cellcolor{green!10} 1.000 &  0.966 \\ 
     -  &    0.75 &   48.423/100 & 5.742 & \cellcolor{red!10} 0.077 &  0.953 \\ \rowcolor{Gray}
     -  &    0.90 &   83.760/100 & 6.346 &  0.000 &  0.157 \\ 
    \hline
\end{tabular}
}
\caption{Robustness to compression from library
  \texttt{keras\_compressor} (SVD-based), compatible with CNN and
  MLP.}
  \label{compression-svd}
\end{table*}

We experimented with a second compression attack: we used a
compression library available on GitHub
(\texttt{keras compressor}\footnote{\url{https://github.com/DwangoMediaVillage/keras_compressor}})
which leverages \textit{singular value decomposition} (SVD) on model
weights, to compress them. We tested it to be compatible with MLP and
CNN. Tab.~\ref{compression-svd} shows that the key extraction on CNN
is not affected by this new attack, as accuracy drops at $88\%$ for
$50\%$ weights compression and with an average of 20 elements removed
(over $42$ tolerated).  The extraction from MLP starts to be affected
with high compression of $75\%$ of the weights, with $48$ elements
removed.

\textit{Overwriting attack via adversarial fine-tuning}\quad
Since we leverage adversaries in the key to embed the watermark in the
model, a plausible attack is to try overwriting this watermark via
adversarial fine-tuning of the leaked model. As explained in Section \ref{ss:fp}, this action also relates
to adversarial training that originally aims to improve model resilience
to adversarial attacks \cite{papernot2017cleverhans}. In this
experiment, we turn $1,000$ images from the MNIST test set into
adversaries and use them to fine-tune the model (the test set being the
  remaining $9,000$ images).  The results of the
overwriting attacks is presented on Tab.~\ref{overwriting}. An
adversarial fine-tuning of size $1,000$ uses $20$ times more
adversaries than the watermarking key (as $|K|=100$, with $50\%$ true
adversaries). We see perfect watermark extractions (no false
negatives) for CNN and MLP, while there are few extraction failures
from the attacked IRNN architecture.  This experiment is thus
consistent with recent arguments about the general difficulty to defend
against adversaries \cite{ensemble,DBLP:journals/corr/abs-1809-02104}.

\begin{table}
  \center
\resizebox{0.5\textwidth}{!}{
\begin{tabular}{|c|ccgc|}
    \hline
    ~    & $K$ elts removed & Stdev &\cellcolor{white!0} Extraction rate & Acc. after \\ \hline
    CNN  & 17.842 & 3.594 &1.000 & 0.983 \\  
    IRNN & 37.423 & 3.931 &0.884 & 0.989 \\
    MLP  & 27.741 & 5.749 &1.000 & 0.972 \\
    \hline
\end{tabular}
}
%\vspace{-0.15cm}
\caption{Robustness to overwriting attacks: Rate of remaining
   zero-bit watermarks in the three attacked models, after model
   fine-tuning with $1,000$ new adversaries.}
\label{overwriting}
\end{table}

  \paragraph{Conclusion on the attacks of watermarks} We highlight that in
  all but two tries, the watermark is robust to considered attacks.  In addition, since
  the attacker cannot know whether her attack is successful or not (as
  the key is unknown to her), each new attack trial is bound to
  degrade the model even further, without removal guarantee. This
  uncertainty will probably considerably discourage trials of this
  kind.

\subsection{About the efficiency and security requirements.} The efficiency requirement deals with the computational cost of querying a suspected
remote service with the $|K|$ queries from the watermarking key.
Given typical pricing of current online machine learning services
(Amazon's Machine Learning, for instance, charges $\$0.10$ per $1,000$
classification requests\ as per Jan. 2018), keys in the order of
hundreds of objects as in our experiments incur financial costs that
are negligible, an indication of negligible computational cost as
well. It is to be noted that if the suspected model is running on an embedded device \cite{8587745}, there is no cost in querying that model; if we target such an application, keys can then be of arbitrary length as far as their embedding preserve the accuracy of the model to be watermarked.
As for the watermarking step (key embedding), it is as complex
as fine-tuning a network using set $K$. Since the size of $K$ is
negligible compared to the original training set size, the
computational overhead of embedding a key is considered low.

%\paragraph{About the security requirement.}
The frontier stitching algorithm deforms slightly and locally the decision frontiers, based on the labelled samples in key $K$. To ensure security, this key must be kept secret by the entity that watermarked the model (otherwise, one might devise a simple overwriting procedure that reverts these deformations). Decision frontier deformation through fine-tuning is a complex process (see work by \cite{DBLP:journals/corr/Berg16}) which seems very difficult to revert in the absence of information on the key (this absence also prevents the use of recent statistical defense techniques~\cite{DBLP:journals/corr/GrosseMP0M17}). Could a method detect specific local frontier configurations that are due to the embedded watermark? The existence of such an algorithm, related to \textit{steganalysis} in the domain of multimedia, would indeed be a challenge for neural network watermarking at large, but seems unlikely.  

 Our watermarking method relies on algorithms for finding
  adversarial examples, in order to get inputs nearby decision
  frontiers. Last few years have witnessed an arms race between attacks
  (algorithms for producing adversarial examples) and defenses to make
  neural networks more robust to those attacks; the problem is still
  open \cite{DBLP:journals/corr/abs-1809-02104} (please refer to
  \cite{adv-survey} for a survey). We argue that our method will
  remain functional regardless of this arms race, as the purpose of
  machine learning for classification is to create decision frontiers
  for separating target classes; there will always be means to cross
  those frontiers, and then to create the inputs we require in order
  to create our watermark keys. Authors in
  \cite{DBLP:journals/corr/abs-1809-02104} precisely characterize
  classes of problems for which adversarial examples cannot be
  avoided, and that depend on properties of the data distribution as
  well as the space dimensionality of the dataset.

\section{Related Work}
\label{sec:related}

Watermarking aims at embedding information into ``objects'' that one
can manipulate locally. One can consider the insertion
  of \textit{visible} watermarks in those objects \cite{visible}; we
  consider in this work and related work the insertion of invisible
  watermarks. Watermarking multimedia content especially is a rich
and active research field, yet showing a two decades old interest
\cite{771066}.  Neural networks are commonly used to insert watermarks
into multimedia content \cite{NN-wat}.

After extension to surprising domains such as network
science~\cite{Zhao:2015:TGW:2817946.2817956}, the extension to
watermarking neural networks as objects themselves is new, following
the need to protect the valuable assets of today's state of the art
machine learning techniques.
Uchida \textit{et al.} \cite{Uchida:2017,Nagai2018} thus propose the watermarking of 
neural networks, by embedding information in the learned weights.
Authors show in the case of convolutional architectures that this embedding does not
significantly change the distribution of parameters in the
model. Mandatory to the use of this approach is a local copy of the neural network
to inspect, as the extraction of the watermark requires reading the
weights of convolution kernels. This approach is motivated
by the voluntary sharing of already trained models, in case of
\textit{transfer learning}, such as in \cite{transfer}'s work for instance.

 Neural network watermarking techniques that allow
  verification in a remote black-box context were recently proposed. These
  works relate to our black box system model (as introduced in the technical report \cite{nous}). 
  They indeed leverage the model's
  outputs to carefully chosen inputs to retrieve the watermark.  
  \cite{zhang2018protecting} proposes to train the
model to be protected with a set of specifically crafted inputs in
order to trigger the assignment of a specific target label by the
model on those inputs. In other words, their approach is very similar
to trojaning \cite{trojaning}: the triggering of a specific label
facing a crafted input constitutes for the authors a proof of
ownership. Similarly, \cite{adi2018turning} explicitly exploits the
possibility of neural network trojaning (\ie, backdooring) for the same
purpose. Finally, \cite{deepsigns} directly embeds the watermark into
the layer weights using specific loss functions. However, contrary to
\cite{Uchida:2017} for instance where the weights directly contain the
watermark (which is therefore not accessible in a black box context),
weights in \cite{deepsigns} are modified in order to produce desired
activations at runtime given specific inputs. Surprisingly, the
authors also show that watermarking mechanisms designed in a  black
box context can also be used in a white box context.
In a more restricted setup, Guo \textit{et al.} propose the adaptation to embedded devices of a black box capable watermark extraction \cite{8587745}.

%Yet, since watermarked objects might only be accessed through API
Since more and more models and algorithms might only be accessed through API
operations (as being run as a component of a remote online service),
there is a growing body of research which is interested in leveraging
the restricted set of operations offered by those APIs to gain
knowledge about the remote system internals. \cite{stealing}
demonstrates that it is possible to extract an indistinguishable copy of
a remotely executed model from some online machine learning APIs.
%while \cite{cscw-nous} propose an algorithm to infer the presence of a
%\textit{centrality} metric in the remote service proposed to users.

  Depth of neural models may also be infered using \textit{timing side channel attacks} \cite{DBLP:journals/corr/abs-1812-11720}.
\cite{Papernot:2017:PBA:3052973.3053009} have shown attacks on remote
models to be feasible, yielding erroneous model outputs.  Authors in
\cite{SETHI2018129} target attacks such as evading a CAPTCHA test, by
the reverse engineering of a remote classifier models.
 Other attacks focus on the stealing of model hyperparameters, from APIs \cite{DBLP:journals/corr/abs-1802-05351,joon2018towards};
\cite{joon2018towards} aims at infering inner hyperparameters (\eg
number of layers, non-linear activation type) of a remote neural
network model by analysing its response patterns to certain inputs.
 Finally, algorithms in \cite{tampernn} propose to detect the
  tampering with a deployed model, also through simple queries to the API; tested attacks are trojaning,
  compression, fine-tuning and the watermarking method proposed in this paper.  In present work, we
propose a watermarking algorithm that is compliant with APIs, since it
solely relies on the basic classification query to the remote service.

\section{Conclusion and perspectives}
\label{discussion}

This article introduces the frontier stitching algorithm, to extract
previously embedded zero-bit watermarks from leaked models that might
be used as part of remote online services. We demonstrated this
  technique on image classifiers; sound \cite{DBLP:journals/corr/abs-1801-01944} and
  video classifiers \cite{DBLP:journals/corr/abs-1807-00458} were also recently found to be prone to adversarial attacks. We believe that a demonstration of existing watermarking techniques in those domains would be of a great practical interest.

%\paragraph{Futurework.}
We focused on classification problems, which account for many
if not most ML-based services. Extensions to other problems (like
regressions or semantic segmentation of images) are a next step for
future work, since adversarial examples also affect those domains.
%prediction models.}

Regarding the model architecture aspect, we have seen that the IRNN model is prone to compression attacks
(pruning rate of $50\%$ of parameters). This underlines the specific
behavior of architectures facing attacks and marking; in depth
 characterization is an interesting future work.

 We challenged the robustness of our watermarking scheme facing
  compression and overwriting attacks. Other more advanced types of attacks might be
  of interest for an attacker that wants to remove the inserted
  watermark. In particular, another attack may be the transfer
  learning of the watermarked model to another task. Recent work
  \cite{transferl} provides a first empirical evidence that the
  adversaries that were integrated in the model through learning
  (defense) might not survive the transfer, leading to a potentially successful
  watermark removal. A full characterization of the
  resilience of adversaries facing transfer learning attacks is
  of great importance for future work.
 
As another future work, we stress that the watermark information
is currently extracted using the binary answers to the query made on
each object in the key: Whether or not this object is classified by
the remote model as expected in the key label.
Leveraging not only
those binary answers, but also the actual classification issued by
the remote model (or even the classification scores), may allow one to
embed more information with the same watermark size.
Another possible improvement may come from the use of the recent concept of universal
adversarial perturbations (\cite{Moosavi-Dezfooli:2017}): they might be
leveraged to build efficient and robust watermarking
algorithms. Indeed, this method generates adversaries that can fool multiple classifiers at once. Relying on such adversaries in an
extension of our framework might give rise to new, improved
watermarking algorithms for neural networks that are queried in a
black box setup.

 Finally, we recall that the watermarking technique we proposed,
  as well as the ones from the related work
  \cite{zhang2018protecting,adi2018turning,8587745,deepsigns}, make
  the assumption that the watermarked model leaked as a bit-level
  copy. Recent attacks on stealing models yet shown the possibility to
  leak a model by approximating it \cite{stealing} through tailored
  queries. A crucial perspective is thus to investigate
  watermark techniques that can resist this alternative type of attacks.

\begin{acknowledgements}
The authors would like to thank the reviewers for their constructive comments.
\end{acknowledgements}

%%%%%%%%%%%%%%%%%%%%%%%%%%%%%%%%%%%%%%%%%%
%\nocite{*}
\bibliographystyle{spmpsci}
\bibliography{biblio}

\end{document}